\documentclass[journal]{IEEEtran}
\usepackage{cite}
\ifCLASSINFOpdf
  \usepackage[pdftex]{graphicx}
\else
\fi
\usepackage{amsmath}
\usepackage{url}

\begin{document}
\title{Highly Integrated Quantum Key Distribution Transmitter Enabled by Silicon Photonics}

\author{Ren-De Liu,
        Yan-Lin Tang,
        Chen-Guang Dong,
        Yan Ma,
        Guo-Qing Liu,
        Zhi-Lin Xie,
        Shuai Li,
        Mi Zou, \\
        Hao Liang,
        Shi-Biao Tang,
        and~Teng-Yun Chen% <-this % stops a space
\thanks{Manuscript received xxx; revised xxx. This work was supported by the Quantum Science and Technology-National Science and Technology Major Project under Grant 2021ZD0300702 and the Anhui Initiative in Quantum Information Technologies. (Ren-De Liu and Yan-Lin Tang contributed equally to this work.)
(Corresponding authors: Shi-Biao Tang and Teng-Yun Chen.)}
\thanks{Ren-De Liu, Chen-Guang Dong, Hao Liang, and Teng-Yun Chen are with the Hefei National Research Center for Physical Sciences at the Microscale and School of Physical Sciences, University of Science and Technology of China, Hefei, Anhui 230026, China.}%
\thanks{Ren-De Liu, Yan-Lin Tang, Yan Ma, Guo-Qing Liu, Zhi-Lin Xie, Shuai Li, and Shi-Biao Tang are with QuantumCTek Corporation Limited, Hefei, Anhui 230088, China (e-mail: shibiao.tang@quantum-info.com).}%
\thanks{Ren-De Liu and Teng-Yun Chen are with the CAS Center for Excellence in Quantum Information and Quantum Physics, University of Science and Technology of China, Hefei, Anhui 230026, China.}%
\thanks{Chen-Guang Dong, Mi Zou, Hao Liang, and Teng-Yun Chen are with the Hefei National Laboratory, University of Science and Technology of China, Hefei, Anhui 230026, China (e-mail: tychen@ustc.edu.cn).}%
}

\maketitle

\begingroup
\footnotesize
\noindent
This is the accepted version of the following article:
``Highly Integrated Quantum Key Distribution Transmitter Enabled
by Silicon Photonics,'' accepted for publication in the
\textit{Journal of Lightwave Technology}.

\noindent
DOI: \url{https://doi.org/10.1109/JLT.2026.3746548}

\medskip
\noindent
\textcopyright\ 2026 IEEE. Personal use of this material is
permitted. Permission from IEEE must be obtained for all other
uses, in any current or future media, including reprinting/
republishing this material for advertising or promotional
purposes, creating new collective works, for resale or
redistribution to servers or lists, or reuse of any copyrighted
component of this work in other works.
\par
\endgroup

\begin{abstract}
Quantum key distribution (QKD) provides information-theoretic security independent of computational assumptions, yet its widespread deployment remains constrained by bulky and costly hardware, with practical transmitters still predominantly implemented as rack-mounted systems. Here, we demonstrate a highly integrated standalone QKD transmitter that incorporates all essential functionalities required for practical QKD operation within a compact platform built around a silicon photonic encoding chip. The transmitter occupies only $167 \times 56 \times 21~\mathrm{mm}^3$ ($\sim0.2~\mathrm{dm}^3$), comparable in size to a half-height, half-width PCIe card and more than 30 times smaller in volume than a conventional 1U rack-mounted system. Paired with a conventional discrete-component receiver, it achieves  secure key rates of 219.1~kbps over 51.3~km and 17.9~kbps over 101.8~km of standard single-mode fiber, with performance comparable to that of conventional discrete-component implementations. This work advances QKD transmitters from conventional rack-mounted equipment to compact board-level platforms, establishing a practical pathway toward module-scale QKD transmitters and broader deployment in quantum-secure communication networks.
\end{abstract}

\begin{IEEEkeywords}
QKD, quantum key distribution, transmitter, silicon photonics.
\end{IEEEkeywords}

% \fi
\IEEEpeerreviewmaketitle

\section{Introduction}
\IEEEPARstart{Q}{uantum} key distribution (QKD) enables information-theoretically secure key exchange and is widely regarded as a promising approach for securing communication in the quantum era~\cite{bernsteinPostquantumCryptography2017,scaraniSecurityPracticalQuantum2009}. Over the past decades, QKD has evolved from proof-of-principle experiments to practical demonstrations in deployed fiber networks~\cite{martinMadQCIHeterogeneousScalable2024,brauerLinkingQKDTestbeds2024,chenImplementation46nodeQuantum2021,chenImplementationCarriergradeQuantum2025} and free-space links, including satellite-to-ground~\cite{liaoSatellitetogroundQuantumKey2017,liMicrosatellitebasedRealtimeQuantum2025} and UAV-based platforms~\cite{tianExperimentalDemonstrationDroneBased2024}, demonstrating its growing maturity for real-world applications. Despite this progress, the widespread deployment of QKD remains constrained by the size, cost, and complexity of practical hardware, which is still predominantly assembled from discrete optical and electronic components in rack-mounted systems.

Addressing this challenge requires substantial hardware miniaturization. The evolution of classical optical communication demonstrates that widespread adoption is closely linked to advances in hardware integration, standardization, and miniaturization, enabling compact, low-cost, and mass-producible transceivers. QKD is expected to follow a similar trajectory. Moreover, QKD offers a unique opportunity for terminal miniaturization: unlike conventional optical communication, it does not require each terminal to support bidirectional user-data exchange; instead, a point-to-point QKD link typically employs separate transmitter and receiver terminals to establish a shared secret key. Consequently, miniaturizing the transmitter alone can provide substantial deployment benefits, particularly for size- and weight-constrained platforms such as UAVs and satellites, and for multi-user access networks, including passive optical networks (PONs), where multiple distributed transmitters can share a centralized receiver.

Driven by this demand, photonic integration has become one of the most active research directions in QKD over the past decade. Silicon-on-insulator (SOI), lithium-niobate-on-insulator (LNOI), indium phosphide (InP), and other integrated photonic platforms have enabled highly integrated quantum-state preparation chips with progressively improved performance and manufacturability~\cite{maSiliconPhotonicTransmitter2016,sibsonChipbasedQuantumKey2017,sibsonIntegratedSiliconPhotonics2017,bunandarMetropolitanQuantumKey2018,gengStableQuantumKey2019,paraisoPhotonicIntegratedQuantum2021,zhangPolarizationBasedQuantumKey2022,liHighrateQuantumKey2023,saxHighspeedIntegratedQKD2023,weiResourceefficientQuantumKey2023,dolphinHybridIntegratedQuantum2023,zhangPolarizationStatesPreparation2025,zhangPolarizationEncodingChips2025,chenIntegratedPhotonicsElectronics2025,linIntegratedLithiumNiobate2025,heoOnchipQuantumKey2025,fan-yuanHighrateQuantumKey2026a}. These advances have significantly increased the integration level of photonic state preparation. However, realizing a deployable standalone QKD transmitter requires far more than photonic state preparation alone. Practical operation additionally requires random-number generation, protocol execution, synchronization, authentication, thermal management, and system control, all of which remain largely implemented using discrete hardware. As a result, most reported systems are still laboratory prototypes or 1U rack-mounted instruments~\cite{paraisoPhotonicIntegratedQuantum2021,chenIntegratedPhotonicsElectronics2025,pereiraQuantumKeyDistribution2026a}, and the miniaturization achieved at the photonic-chip level has not yet been fully translated into compact standalone QKD transmitters. Advancing complete QKD transmitters toward board- and ultimately module-level platforms therefore represents an important step toward practical deployment.

\begin{figure*}[!t]
\centering
\includegraphics[width=\textwidth]{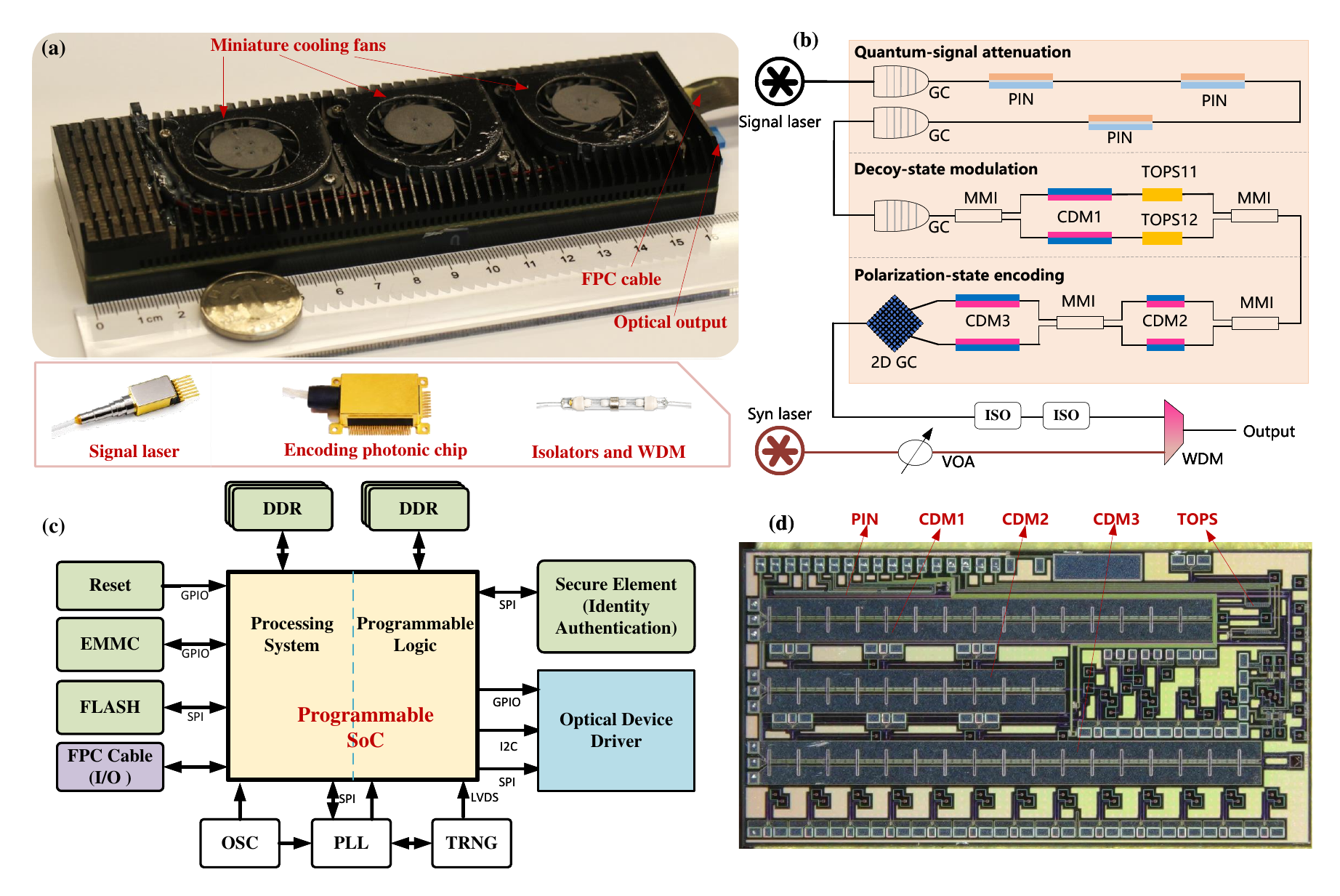}
\caption{(a) Photograph of the QKD transmitter; (b) schematic of the optical subsystem which comprises a silicon photonic chip together with compact packaged optical components. The silicon photonic chip contains grating couplers (GCs), carrier-depletion modulators (CDMs), thermo-optic phase shifters (TOPSs), and multimode interferometers (MMIs). The compact packaged optical components include optical isolators (ISOs), a wavelength-division multiplexer (WDM), and a variable optical attenuator (VOA). The pink regions indicate the functions implemented on the silicon photonic chip; (c) block diagram of the electronic subsystem, including a programmable system-on-chip (SoC) and a true random number generator (TRNG); and (d) microscope image of the encoding silicon photonic chip.}
\label{fig1-tx}
\end{figure*}

Here, we demonstrate a highly integrated standalone QKD transmitter built around a silicon photonic encoding chip fabricated using a commercial SOI process. The transmitter incorporates all essential functionalities required for practical QKD operation within a unified board-level architecture. The complete transmitter occupies only $167 \times 56 \times 21~\mathrm{mm}^3$ ($\sim0.2~\mathrm{dm}^3$), comparable in size to a half-height, half-width PCIe card and more than 30 times smaller than a conventional 1U rack-mounted implementation. Paired with a conventional discrete-component receiver, the transmitter achieves secure key rates of 219.1~kbps over 51.3~km and 17.9~kbps over 101.8~km of standard single-mode fiber, demonstrating performance comparable to conventional discrete-component systems~\cite{tangTimebinPhaseencodingQuantum2023,pincemin400GbpsCoherentTransmission2024}. Rather than advancing photonic integration alone, this work establishes system-level integration as the next stage in the evolution of integrated QKD transmitters.

\section{QKD Transmitter Design}
\label{sec:tx_design}
\subsection{Overview}
The fabricated standalone QKD transmitter is shown in Fig.~\ref{fig1-tx}(a). The module incorporates all essential functionalities of a practical QKD transmitter, including optical pulse generation, quantum-signal attenuation, decoy-state modulation, polarization-state encoding, Trojan-horse-attack protection~\cite{lucamariniPracticalSecurityBounds2015}, high-speed true-random-number generation, system control and protocol processing, and identity authentication. It also incorporates a dedicated TO-packaged DFB synchronization laser and a MEMS-based variable optical attenuator (VOA) for clock-synchronization with the receiver, enabling flexible adaptation to different link-loss conditions. To enable reliable long-term operation, the transmitter also incorporates a dedicated thermal management assembly comprising a heat sink and three embedded cooling fans, which account for a substantial fraction of the module volume. To realize these functionalities, the transmitter is organized into tightly integrated optical and electronic subsystems. Quantum-signal attenuation, decoy-state modulation, and polarization-state encoding are implemented on the silicon photonic chip, while the signal laser, synchronization laser, optical isolators, WDM, and synchronization-path VOA are realized using compact packaged optical components. The protocol-processing, random-number-generation, authentication, control, and driving functions are implemented at the board level around a programmable SoC. The detailed implementations of these subsystems are described in the following sections.

\subsection{Optical Design}
The optical subsystem of the transmitter, illustrated in Fig.~\ref{fig1-tx}(b), consists of a quantum signal path for quantum-state preparation and a synchronization path for receiver clock synchronization. The subsystem follows a hybrid integration strategy, in which the high-speed quantum-state preparation functions are monolithically integrated on a silicon photonic chip, while the remaining optical functions are realized using miniaturized optical components and compact packaged devices to achieve a highly integrated yet practical optical implementation.

\textbf{Signal pulse generation:}
Signal pulses are generated by a miniaturized 9-pin box-packaged distributed feedback (DFB) laser operating in gain-switched mode with zero DC bias current, facilitating pulse-to-pulse global phase randomization~\cite{kobayashiEvaluationPhaseRandomness2014}. The vacuum decoy state is generated by direct modulation of the signal laser.

\textbf{Quantum-state preparation on the silicon photonic chip:}
Following pulse generation, the optical signal is coupled into a thermally stabilized, box-packaged silicon photonic chip. Fabricated using a standard SOI multi-project wafer (MPW) process, the chip integrates quantum-signal attenuation, decoy-state modulation, and polarization-state encoding. As highlighted in pink in Fig.~\ref{fig1-tx}(b), these functions are monolithically integrated on the silicon photonic chip, whose microscope image is shown in Fig.~\ref{fig1-tx}(d). The optical-attenuation and intensity-modulation sections are interconnected through two grating couplers and an external fiber connection, while the intensity-modulation and polarization-encoding sections are directly connected by on-chip waveguides.

The input pulse is coupled into the chip through a grating coupler. It first passes through a three-stage PIN-based VOA, which precisely controls the mean photon number in the quantum regime. The pulse then enters a high-speed decoy-state modulation stage comprising two multimode interferometers (MMIs), two thermo-optic phase shifters (TOPS11/12), and a carrier-depletion modulator (CDM1). Subsequently, a power-balancing stage, consisting of another pair of MMIs followed by CDM2, adjusts the relative optical power between the two waveguide components. Finally, high-speed polarization-state encoding is realized by CDM3 and the two-dimensional grating coupler. The two TE-mode components propagate through the two waveguide arms to the 2D grating coupler, where they are converted into two orthogonal polarization components. With the relative optical powers preset by CDM2, CDM3 introduces four discrete relative phase shifts between the two components, thereby generating the four polarization states required for the BB84 protocol.

\textbf{Trojan-horse attack protection:}
Trojan-horse attacks represent an important practical security consideration for QKD transmitters~\cite{lucamariniPracticalSecurityBounds2015,jainRiskAnalysisTrojanHorse2015,ponosovaProtectingFiberOpticQuantum2022}. Trojan-horse attack protection is implemented through a combination of optical isolation, wavelength-selective filtering, and coupling loss associated with the silicon photonic interface. Two standalone isolators are placed at the transmitter output, while the WDM incorporates an additional optical isolator and simultaneously combines the quantum signal and the synchronization light. For Trojan-horse light injected from the channel side, broadband characterization of the implemented protection path shows an estimated total attenuation exceeding 170~dB over 1250--1650~nm. Previous security analysis has shown that approximately 170~dB of total optical isolation can substantially reduce the impact of Trojan-horse leakage on the secret key rate under the corresponding assumptions~\cite{lucamariniPracticalSecurityBounds2015}. This configuration provides substantial engineering mitigation of Trojan-horse probing while maintaining a compact optical layout.

\textbf{Synchronization light path:}
Receiver synchronization is realized using a dedicated TO-packaged DFB laser. A compact MEMS-based VOA, with an insertion loss below 1~dB, enables flexible adaptation to different fiber link losses, ensuring reliable clock synchronization over various transmission distances.

Together, these components realize a hybrid optical architecture that combines monolithic silicon photonic integration with compact packaged optical components, enabling a highly integrated optical subsystem while preserving the functionality required for practical standalone QKD transmitters.

\subsection{Electronics Design}
As illustrated in Fig.~\ref{fig1-tx}(c), the electronic subsystem adopts a highly integrated architecture centered on a programmable system-on-chip (SoC). Unlike conventional QKD transmitters that typically employ separate CPU and FPGA devices, the proposed design integrates processor cores and programmable logic within a single SoC, significantly reducing board area, interconnection complexity, and power consumption while maintaining flexible real-time protocol execution and precise timing control.

The SoC is supported by peripheral memory modules, including DDR, eMMC, and FLASH, which provide data buffering and storage for protocol execution. Stable system timing is ensured by an on-board oscillator together with a phase-locked loop (PLL).

A dedicated hardware true random number generator (TRNG, QuantumCTek QCTWNG)~\cite{wang10GbpsTrueRandom2016} supplies high-speed random numbers directly to the SoC through a low-voltage differential signaling (LVDS) interface. These random bits are used for decoy-state modulation, basis selection, and bit generation in the BB84 protocol.

High-speed digital signals generated by the SoC are serialized and routed to the photonic chip driver circuitry. For decoy state modulation, CDM1 is directly driven by a MAX3942 10-Gbps modulator driver without an additional RF amplifier. After fine adjustment of the operating bias point, the modulation swing used for decoy-state preparation is approximately 1.5~V${\mathrm{pp}}$. For polarization encoding, the outputs of three MAX3942 modulator drivers are combined through a resistive network and subsequently amplified by a commercial HMC659 RF amplifier to generate the four-level electrical drive required by CDM3. The available output swing is approximately 5.9~V${\mathrm{pp}}$, while the polarization state requiring the largest drive corresponds to an applied voltage of approximately 5.3~V. Low-speed control signals are generated by conventional digital-to-analog converters (DACs) and amplifiers to drive slow-response components such as bias circuits and VOAs, with configuration managed through an I\textsuperscript{2}C interface. Identity authentication and key protection are provided by a dedicated secure element connected via the SPI interface. The transmitter communicates with external systems through a flexible electrical interface, enabling compact and reliable system integration.

The highly integrated SoC architecture, together with the dedicated hardware TRNG, enables real-time execution of the BB84 protocol within a compact electronic subsystem. Experimental validation of protocol operation and secret-key generation is presented in the following section.

\section{QKD Experiments}
\subsection{Experimental Setup}
\begin{figure}[!t]
\centering
\includegraphics[width=\columnwidth]{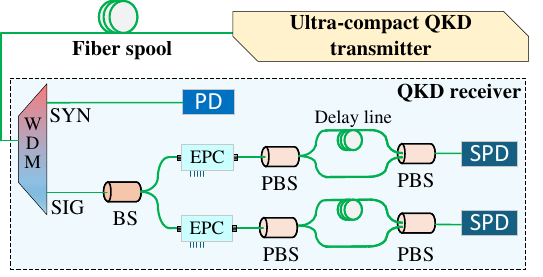}
\caption{QKD experimental setup. Optical signals from the proposed standalone QKD transmitter are delivered to the receiver via a standard fiber spool. At the receiver, the quantum signal (SIG) and synchronization signal (SYN) are demultiplexed by a WDM. A 50:50 beam splitter (BS) performs passive basis selection, and electrical polarization controllers (EPCs) compensate for polarization fluctuations in the transmission fiber. Quantum states in the same basis are measured using polarization beam splitters (PBSs), time-division multiplexed by a PBS and delay lines, and finally detected by single-photon detectors (SPDs).}
\label{fig2-setup}
\end{figure}

The experimental setup is illustrated in Fig.~\ref{fig2-setup}. The transmitter is the proposed highly integrated standalone QKD transmitter described in Section~\ref{sec:tx_design}, while the receiver is implemented using conventional discrete optical components. This configuration allows the performance of the proposed transmitter to be evaluated independently of receiver-side integration. The system implements the standard decoy-state BB84 protocol~\cite{hwangQuantumKeyDistribution2003,wangBeatingPhotonNumberSplittingAttack2005,loDecoyStateQuantum2005,maPracticalDecoyState2005}.

The transmitter operates at a pulse repetition rate of 625~MHz with an optical pulse width (FWHM) of 41.6~ps at a wavelength of 1550.12~nm. Quantum states are prepared in four polarization states forming two mutually unbiased bases (Z and X), with each state selected with equal probability. A three-intensity decoy-state scheme is adopted, using mean photon numbers of $\mu=0.3$, $\nu=0.1$, and $w<0.003$ for the signal, weak-decoy, and vacuum states, respectively, with corresponding probabilities of $6/8$, $1/8$, and $1/8$. A dedicated synchronization channel operating at 1569.59~nm and 100~kHz provides the timing reference between the transmitter and receiver.

At the receiver, the incoming optical signals are first demultiplexed into the synchronization and quantum channels by a WDM. Passive basis selection is implemented using a 50:50 beam splitter (BS), while electrical polarization controllers (EPCs) compensate for polarization drift introduced by the transmission fiber. Within each basis, the orthogonal polarization components are first separated by a polarization beam splitter (PBS). After one polarization component is delayed, the two components are recombined by a second PBS to form a time-division multiplexed (TDM) signal, allowing both polarization states to be detected by a single InGaAs single-photon detector (SPD). The two SPDs operate at a repetition rate of 1.25~GHz, each with a detection efficiency of approximately 25\%, a dark-count rate below 800~cps, and a dead time of 200~ns.

Compared with the conventional four-detector architecture widely adopted in laboratory QKD systems, this two-detector receiver substantially reduces receiver complexity, cost, and detector count while providing a more practical architecture for future receiver miniaturization and integration. Conventional coupler-based TDM implementations introduce an inherent multiplexing loss of approximately 3~dB. By employing the PBS-based TDM scheme, the multiplexing loss is reduced to below 0.5~dB. Consequently, the total additional optical loss introduced by the receiver is kept below 2~dB.

Despite this optimization, the two-detector TDM architecture still incurs additional optical loss and increased detector afterpulsing compared with a conventional four-detector receiver, leading to a lower achievable secure key rate. Accordingly, the secure key rates reported in this work should be regarded as a conservative evaluation of the proposed transmitter under a deployment-oriented receiver architecture that is better suited for future miniaturization and low-cost deployment, rather than the maximum performance achievable with a conventional four-detector laboratory receiver.

\subsection{Real-Time Protocol Processing}

Beyond quantum-state generation, the proposed transmitter integrates a complete real-time QKD protocol-processing framework within the embedded SoC, including message authentication, basis sifting, error correction, error verification, and privacy amplification for the standard decoy-state BB84 protocol. Unlike conventional laboratory implementations that rely on external computers for protocol execution, all protocol-processing tasks are executed locally within the transmitter, enabling real-time secure-key generation in a compact standalone platform.

Message authentication is initialized using pre-injected secret keys securely stored in the on-board secure element. Consumed authentication keys are immediately replaced with newly generated quantum keys, ensuring continuous authenticated operation. When the remaining authentication-key pool falls below a predefined threshold, the system automatically generates a warning for key replenishment.

\begin{figure*}[!t]
\centering
\includegraphics[width=\textwidth]{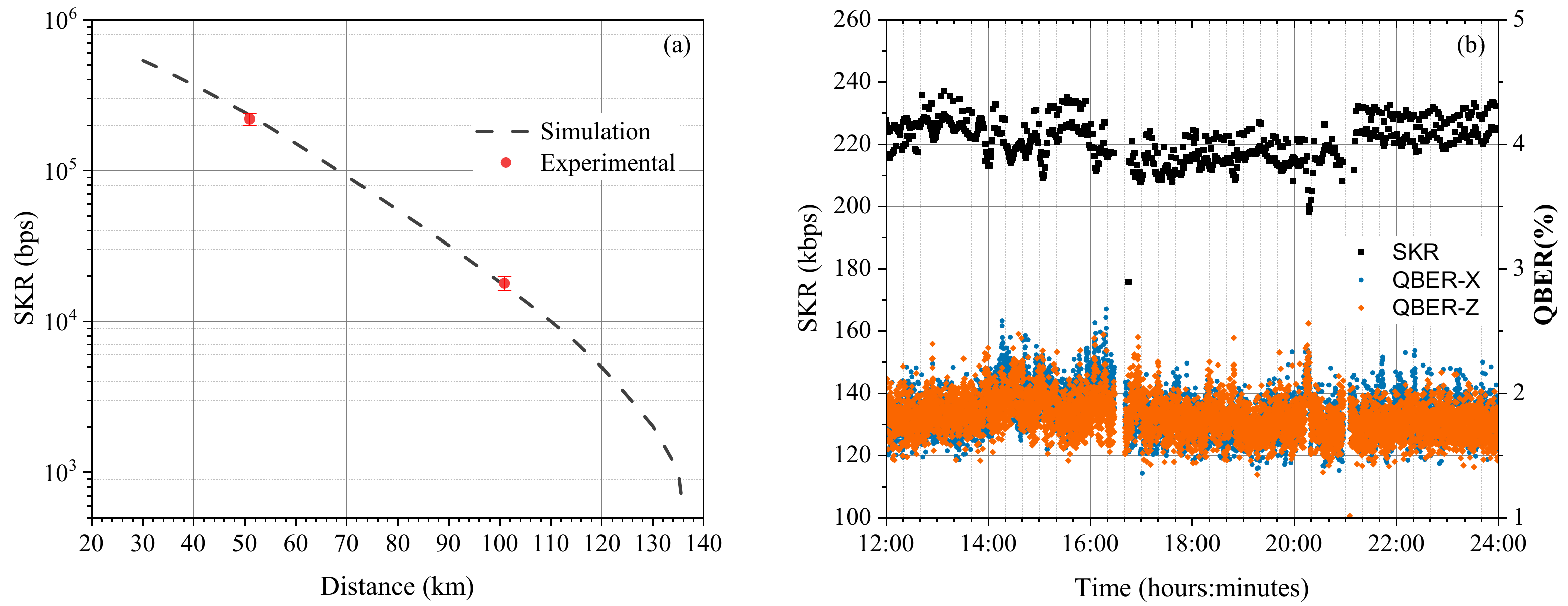}
\caption{(a) Simulated secret-key rate (SKR) as a function of transmission distance, together with experimental results at 51.3 km and 101.8 km; (b) SKR and QBERs of the X and Z bases measured during a 12-hour continuous operation over 51.3~km of standard single-mode fiber.}
\label{fig3-SKR}
\end{figure*}

\begin{table*}[!t]
\caption{Comparison of representative integrated QKD transmitter implementations.}
\label{tab:qkd_transmitter_comparison}
\centering
\small
\renewcommand{\arraystretch}{1.12}
\setlength{\tabcolsep}{6pt}

\begin{tabular}{lccccc}
\hline
\textbf{Reference} &
\textbf{Platform} &
\textbf{Security functions} &
\textbf{Real-time processing} &
\textbf{Thermal management} &
\textbf{Form / Volume} \\
\hline

Ma~\cite{maSiliconPhotonicTransmitter2016}
& SOI PIC
& NR
& Ext.
& NR
& Laboratory \\

Sibson~\cite{sibsonChipbasedQuantumKey2017}
& InP PIC
& NR
& Ext.
& NR
& Laboratory \\

Para\"iso~\cite{paraisoPhotonicIntegratedQuantum2021}
& InP PIC
& NR
& Int.
& Int.
& 1U rack \\

Chen~\cite{chenIntegratedPhotonicsElectronics2025}
& SOI PIC + EIC
& NR
& Ext.
& NR
& Laboratory \\

Pereira~\cite{pereiraQuantumKeyDistribution2026a}
& SOI PIC + EIC
& NR
& Int.
& Int.
& 1U rack \\

\textbf{This work}
& \textbf{SOI PIC}
& \textbf{Auth. + THP}
& \textbf{Int.}
& \textbf{Int.}
& \textbf{Board-level (~0.20 dm$^3$)} \\

\hline
\end{tabular}

\vspace{1mm}

\scriptsize
NR, not reported; Auth., authentication; THP, Trojan-horse protection;
Ext., external/offline; Int., integrated; EIC, electronic integrated circuit.
\end{table*}

Following quantum transmission, detection events received from the receiver are processed in real time by the embedded FPGA. Basis sifting is followed by error correction using the Winnow algorithm and error verification using CRC-64. Privacy amplification based on Toeplitz hashing is then applied to 6-Mbit blocks of error-corrected and CRC-verified key. For the 51.3-km link, the accumulation time for one such block is approximately 11~s. The privacy amplification process removes residual information potentially available to an eavesdropper, yielding the final information-theoretically secure key.

By integrating the complete post-processing chain within the embedded SoC, the proposed transmitter performs autonomous protocol execution without external computing hardware. Together with the integrated optical subsystem described in  Section \ref{sec:tx_design}, the platform constitutes a complete standalone QKD transmitter rather than merely an integrated photonic state-preparation module.

\subsection{System-Level Demonstration}
The proposed standalone QKD transmitter was experimentally evaluated over standard single-mode fiber under the composable finite-size security framework against general attacks~\cite{fungPracticalIssuesQuantumkeydistribution2010,tomamichelTightFinitekeyAnalysis2012,limConciseSecurityBounds2014,zhangImprovedKeyrateBounds2017,tangFieldTestQuantum2023}. The secure key length is calculated as

\begin{equation}
L = M_{1zz}^{L}\left[1-H_2(e_{1zz}^{pU})\right]
  + M_{1xx}^{L}\left[1-H_2(e_{1xx}^{pU})\right]
  -\mathrm{leak}_{\mathrm{EC}},
\end{equation}
where $M_{1zz}^{L}$ ($M_{1xx}^{L}$) is the lower bound of single-photon event counts of the Z (X) basis after basis sifting in the signal state. $e_{1zz}^{pU}$ ($e_{1xx}^{pU}$) is the upper bound of the single-photon phase error rate of the Z (X) basis after basis sifting in the signal state, detailed calculated following~\cite{tangFieldTestQuantum2023}. $H_2(x)$ denotes the binary entropy function, and $\text{leak}_{\text{EC}}$ represents the information leakage during error correction.

Figure~\ref{fig3-SKR}(a) presents the measured secure key rate (SKR) together with the theoretical prediction as a function of transmission distance. Secure key rates of 219.1~kbps and 17.9~kbps are achieved over 51.3~km (10.18~dB loss) and 101.8~km (20.17~dB loss) of standard single-mode fiber, respectively. The excellent agreement between experiment and theory confirms the correct operation of the integrated optical subsystem, embedded electronics, and real-time protocol-processing framework, demonstrating that the proposed standalone transmitter performs reliably over a wide range of transmission distances.

Figure~\ref{fig3-SKR}(b) further demonstrates the long-term operational stability of the proposed transmitter. During a continuous 12-hour experiment over 51.3~km of standard single-mode fiber, the system maintained stable secure-key generation, with the secure key rate and the QBERs of both the X and Z bases exhibiting only minor fluctuations throughout the measurement period. These results confirm the robustness and stability of the integrated hardware and embedded protocol-processing framework during extended operation. Under normal QKD operation, the measured total electrical power consumption of the transmitter is approximately 25~W.

While the above experiments verify the transmission performance of the proposed system, the principal contribution of this work lies in the realization of a highly integrated standalone QKD transmitter. Representative integrated QKD transmitter implementations are therefore compared in Table~\ref{tab:qkd_transmitter_comparison}.

Previous studies have primarily focused on photonic integration or photonic--electronic integration, with most implementations remaining laboratory prototypes or rack-mounted instruments. In contrast, the proposed transmitter consolidates complete QKD functionality into a compact board-level platform, reducing the transmitter volume by more than a factor of 30 compared with a conventional 1U implementation while maintaining practical QKD performance.

\section{Conclusion and Outlook}
In this work, we have demonstrated a highly integrated standalone QKD transmitter that combines silicon photonics with embedded electronics, real-time protocol processing, authentication, and thermal management in a compact board-level platform. Experimental results show that substantial system-level miniaturization can be achieved while maintaining practical QKD performance over metropolitan-scale fiber links. These results demonstrate the feasibility of advancing complete QKD transmitters from conventional rack-mounted equipment toward compact board-level platforms without compromising practical system performance.

The proposed architecture provides a practical pathway toward compact, scalable, and cost-effective QKD transmitters for applications ranging from size- and weight-constrained platforms, such as UAVs and small satellites, to large-scale access networks, including PONs, where multiple transmitters can share a centralized receiver. More broadly, this transition toward compact board-level transmitters follows the miniaturization trajectory of classical optical communication and represents an important step toward broader deployment of quantum-secure communication infrastructure.

Future work will focus on further heterogeneous integration of photonic and electronic subsystems, receiver miniaturization, and higher levels of system integration. Leveraging mature commercial silicon photonic technology, the proposed architecture also offers a promising route toward scalable manufacturing and practical deployment of integrated QKD systems.

\section*{Acknowledgment}

The authors would like to thank J-P.\ Xu for helpful discussions and Y-L.\ Li for assistance with the experiments.

\ifCLASSOPTIONcaptionsoff
  \newpage
\fi

\bibliographystyle{IEEEtran}
\bibliography{bibtex/bib/IEEEabrv,bibtex/bib/references}

\end{document}